\documentclass[aps,pra,twocolumn,superscriptaddress,10pt]{revtex4-2}
\usepackage{graphicx}
\usepackage{xcolor, soul}
\usepackage{hyperref}
\usepackage{amsmath}
\usepackage{amssymb}
\begin{document}

\title{Highly correlated electronic bounding and spin effect: \\ confirmation of an autodetaching state of O$^-$.}

\author{M. M. Sant'Anna}
\affiliation{Instituto de F\'isica, Universidade~Federal~do~Rio~de~Janeiro,
             Caixa Postal 68528, Rio de Janeiro, - RJ 21941-972, Brazil.}
             
\author{Mart\'{i}nez-Calder\'{o}n A. A.}
\email{amartinez@icf.unam.mx}
\affiliation{Instituto de Ciencias F\'{i}sicas, Universidad Nacional Aut\'onoma 
             de México, Avenida Universidad s/n, Col. Chamilpa, Cuernavaca, Morelos, 
             62210 Mexico.}

\email{mms@if.ufrj.br}
\author{Ginette Jalbert}
\affiliation{Instituto de F\'isica, Universidade~Federal~do~Rio~de~Janeiro,
             Caixa Postal 68528, Rio de Janeiro, - RJ 21941-972, Brazil.}
\email{ginette@if.ufrj.br}
\author{A. B. Rocha}
\email{rocha@iq.ufrj.br}
\affiliation{Instituto de Qu\'{i}mica, Universidade~Federal~do~Rio~de~Janeiro,
             Caixa Postal 68528, Rio de Janeiro, - RJ 21941-972, Brazil.}

\author{G. Hinojosa} 
\email{hinojosa@icf.unam.mx}
\affiliation{Instituto de Ciencias F\'{i}sicas, Universidad Nacional Aut\'onoma 
             de México, Avenida Universidad s/n, Col. Chamilpa, Cuernavaca, Morelos, 
             62210 Mexico.}

\date{\today}

\begin{abstract}

{\bf Reproduced from: Phys. Rev. A,  Vol. {\bf 112}, 052823, Published Nov., 20$^{th}$ 2025, with the permission of AIP Publishing. 
This article may be downloaded for personal use only. Any other use requires prior permission AIP Publishing. 
The article may be found at: DOI: https://doi.org/10.1103/2st9-v5h1. } \ \\

The existence of an auto-detaching state of O$^-$ with a lifetime on the scale
of a hundred nanoseconds is demonstrated both experimentally and theoretically.
The O$^-$ lifetime values are determined using two recently developed methods.
The experimental approach is based on a derivation from measured electron-loss 
cross sections combined with time-of-flight spectrometry. For the theoretical 
approach, the continued Green's function within the formalism of Fano-Feshbach is applied.
We present the measured lifetime value of $100 \pm 10 \text{ ns}$.
The calculated lifetime value is 75 ns, and is associated with the (2p$^3$3s$^2$)$^4$S state of O$^-$.
We discuss how the existence of a 100-ns-lifetime oxygen metastable anion can impact 
the modeling of oxygen-containing systems.   

\end{abstract}


\maketitle 

The electronic structure of negative ions is dominated by electronic correlation 
effects \cite{Kristiansson2022,andersen2004,pegg2004}. Unlike atoms and positive ions, 
anions present 
bound-state spectra with a finite number of energy levels. In this respect, they are more 
alike to nuclei than to atoms or positive ions. The search and detection of anions 
in space and in Earth's atmosphere, especially H$^-$ and O$^-$, triggered the early 
development of theoretical methods for studying their electronic structure, in 
configurations ranging from the ground state and possible singly-excited states 
\cite{chandrasekhar1944,chandrasekhar1944b,bates1943,massey1942} to doubly-excited 
self-detaching states \cite{rau1992,rau1996}. While for H$^-$ some attempts to measure 
the lifetimes of auto-detaching states have been made \cite{berry1975,balling2000, vergara2021}, 
for O$^-$, there are no reports of similar efforts.

The interaction between slow anions and atomic or molecular partners, either neutral
or charged, gives rise to the so-called chemistry of negative ions \cite{desai2021}. 
The creation and destruction of O$^-$ in a chemical environment are related to
several processes, including atomic collisions, dissociative electron attachment
(DEA) to molecules \cite{nandi2005,pshenichnyuk2022,xie2024}, and photodetachment 
\cite{pegg2023}. These interaction processes occur within different time scales. 
The analysis of the role played by excited states in these systems depends on the 
knowledge of the particular intrinsic time scale of the state decay. 
Therefore, incomplete knowledge of the auto-detaching states of O$^-$ precludes a 
careful evaluation of their role in time-dependent physical systems for which 
electronic excitation is possible.

On Earth, O$^-$ is present in several physical systems, such as in earth's troposphere 
\cite{chick2023}, in flames \cite{ayasli2024,kim2015,popov2022}, and in 
sputtering plasmas \cite{depla2024}. Extraterrestrial planetary environments may involve 
O$^-$, like in the lower ionosphere of Mars \cite{molina2002} and in the ionosphere 
of Titan \cite{mukundan2018,nixon2024}. In the latter case, O$^-$ is described by 
Mukundan and Bhardwaj as a 
transient formed from photofragmentation of H$_2$O, reacting immediately with 
CH$_4$ or HCN to produce 
molecular anions, although they do not quantitatively discuss this immediacy \cite{mukundan2018}. 
Ayasli {\it et al.} \cite{ayasli2024} have recently observed long-lived transition states in 
low-energy O$^-$ + CH$_4$ collisions. The presence of O$^-$ long-lived excited states
can influence reaction rates, affecting the genesis of molecular anions formed as byproducts
of such transition states.


The K-shell photoexcitation of small molecules also produces O$^-$ fragments. The 
photon-energy dependencies of the produced O$^-$ through photofragmentation of CO and 
H$_2$O show signatures of post-collision interaction  \cite{hansen2002, stolte2003}. 
Linewidth measurements with CO target provide both, core-hole and vibrational of the order 
of 10 fs  \cite{hansen2002}. 
The post-collision interaction characteristic time, however, is not measured or 
calculated. The electron recapture process is inherent to PCI. If auto-ionizing excited 
states of O$^-$ live long enough, they may influence the recapture dynamics. This example 
shows that measurements of lifetimes for O$^-$ excited states are important not only 
for providing time scales but also for helping to choose which excited states should be 
considered in the theoretical modeling of physical systems producing O$^-$ ions. 

Lifetimes for anionic excited states are essential for testing time-dependent 
quantum calculation methods and for estimating the impact of excited states in relevant physical 
systems. 

To our knowledge, there exists a lifetime measurement for the O$^-$ K-shell excited 
1s$^1$2s$^2$2p$^6$ state \cite{schippers2020}. There are also lifetime calculations for 
the (1s$^2$2s$^2$2p$^5$)$^2$P$_{1/2}$ state, which is different from the ground state 
(1s$^2$2s$^2$2p$^5$)$^2$P$_{3/2}$ only in the fine structure (with $\Delta$E= 0.022 eV 
\cite{Kristiansson2022}), estimating a lifetime of the order of hours for that metastable 
state \cite{brage2017}. For valence shell states, we could not find in the literature 
any lifetime measurement values for the excited states of O$^-$.

The existence of doubly excited autodetaching states of O$^-$ has been known for decades
\cite{esaulov2017}. The spectroscopy of these states is known from the 
detection os emitted photons or electrons in collisions involving the O$^-$ projectile and 
several targets \cite{edwards1971}. 
However, those measurements do not provide lifetime values, since linewidth-based measurements 
in general are limited by the instrumental resolution (e.g. \cite{schlachter2004}). 
In this letter we present a lifetime measurement for an O$^-$ doubly excited state,
obtained using a novel technique based on the O$^-$ projectile time of flight. We also present an
independent Fano–Feshbach formalism calculation that includes an analytic continuation.



The present experimental method has been described previously in some detail \cite{AAmartinez2024} 
in a derivation for the lifetime of a possible auto-detaching state of H$^-$ \cite{vergara2021}. 
When the present Eq. (\ref{betapar}) for $^m\sigma$ is applied to the previous analysis for H$^-$, 
a new value for the auto-detaching state of $25 \text{ ns} \pm \text{3 ns}$ is derived, 
which we compared with the only available value for $\tau$ of 17 ns by Drake 
\cite{drake1973} for the 2p$^2$($^3$P) of H$^-$. This particular state of H$^-$ has 
been investigated before by Bunge and collaborators \cite{bunge1982} and by Bylicki 
and Bednarz \cite{bylicki2003}. Single excited states of H$^-$ are short-lived (of the 
order of 10$^{-14}$ s \cite{corderman1979}), and are unresolved in the present experiment.

\section{Theory}
In a 
previous paper \cite{oliveira2022fano}, two of us have developed a method to calculate 
the autoionization width from a discretized pseudospectrum. The method is based on the 
analytic continuation of the Green’s function within the Fano-Feshbach formalism. This
approach is commonly used in quantum mechanics and many-body theory to study complex 
systems and their behavior. The Green’s function provides valuable insights into the 
density of states and other physical quantities. By analytically continuing the Green’s 
function, we explore its behavior in different regions of the complex plane, revealing 
important features of the system. In brief, a pseudo spectrum is constructed from methods 
relying on  a $L^2$ square-integrable basis set. The energy is extended to the complex 
plane and the analytical continuation is performed through Padé approximants or, 
equivalently, via continued fractions. The ionization width is obtained from the 
imaginary part of the complex energy.

In the present case, the same method is used to obtain the autodetachment lifetime. 
For O$^-$, the pseudo-spectrum is calculated as follows. Initially, a full valence 
CASSCF method was used to obtain the lowest-lying doublet and quartet states. 
Then, these states are used as basis to diagonilize the full Breit-Pauli Hamiltonian 
in order to obtain the spin-orbit manifold, which is finally used as pseudo-spectrum. 
The corresponding energies are extended to the complex plane where the analytical 
continuation is performed. The basis set used in all calculations is aug-cc-pVTZ 
complemented by Kaufmann’s continuum-like basis set. These continuum-like functions 
are essential to properly describe the pseudo-continuum part of the spectrum and is 
explicitly presented in ref \cite{Tenorio18}.

\section{Experiment}

Here, we offer the advancement of the new method to derive lifetimes ($\tau$) based on 
collision-induced total electron loss cross-sections (CS) measurement of the anionic 
projectile O$^-$ on O$_2$ and on N$_2$ as a function of the projectile kinetic energy. 
To our knowledge, there exists only one previous experimental approach to derive lifetimes 
from cross-sections, in that case, from dissociative electron detachment cross-sections 
for diatomic-like molecules \cite{marton1978}. 

The present method consists of the application of two different well-established techniques 
for the measurement of electron loss cross sections, namely, the beam attenuation (BAT) 
and the signal growth rate (SGR). BAT is based on measuring the remaining negative
ions from an O$^-$ ion beam after interacting with a target gas. SGR consists of
measuring the resulting neutral atoms of oxygen produced from the electron loss
after the interaction.

In the following, a brief description of the experimental method is given. Negative ions 
were extracted from a quartz-made chamber \cite{colutron}, containing a mixture of 
high-purity O$_2$ and Ar gases, where a plasma was induced by an incandescent filament. 
The ion source chamber was operated at a maximum pressure of 20 Pa and, had an exit orifice 
of 1.0 mm diameter on a 100 V biased metallic cap. The negative ions were expelled toward 
a set of electrostatic lenses that focused them in a magnetic field region where the 
O$^-$ ions were separated according to their mass-to-charge ratio. 

 \begin{figure}
 \begin{center}
  \includegraphics[width=\columnwidth]{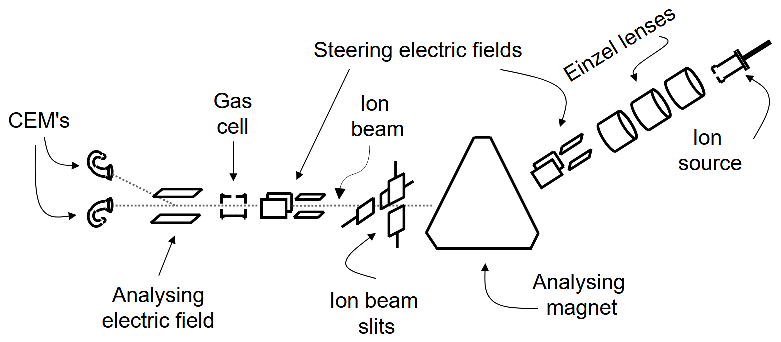}
   \caption{\label{machine} Schematic diagram of the apparatus. Not to scale.}
 \end{center}  
 \end{figure} 
 
 Before and after the magnetic field's region, two sets of parallel plates installed 
 perpendicularly to the ion-beam direction, were used to apply  small electric fields 
 to fine-tune the ion beam trajectory.
 
 A well-collimated, single-species, monoenergetic, and, focused ion beam, with a width
 of $\approx$ 0.5 mm, enters a gas cell where the interaction of O$^-$ and O$_2$ 
 occurs in a time frame of a fraction of 10$^{-15}$ s,

\begin{equation}\label{collisEq}
 \mbox{O}^- + \mbox{O}_2 \longrightarrow \mbox{O} + \mbox{e}^- + \mbox{O}_2^*
\end{equation} 
 
 \noindent where the star is to indicate an unknown final internal state. High-purity 
 O$_2$ gas was used. The resulting neutral oxygen atoms continued in their original trajectory 
 towards a channel electron multiplier (CEM) located in the symmetry axis of the apparatus.
 The remaining O$^-$ ions from the parent ion beam were diverted by a perpendicular electric 
 field set (see Fig. \ref{machine}) to steer the ions toward a second CEM installed 
 off-axis of the accelerator symmetry axis. This Analyzing Electric Field will be referred to 
 as AEF.
 
 The transit time of a 1.0 keV O$^-$ ion from the output of the ion source to the entrance of 
 the gas cell was approximately 7 $\mu$s. The experiment was performed under high vacuum 
 conditions so that the mean free path of the ion beam was larger than its total trajectory, 
 except inside the gas cell.
 
 To verify the full collection of the charged particles, the ion-beam intensity was 
 measured as a function of the AEF intensity. The resulting profiles showed a plateau,
 indicating that their spatial spread was smaller than the width of the CEMs 
 aperture after dispersion in the gas cell, therefore showing complete collection.

 To check for the full collection of neutral atoms, we implemented a test with the lighter 
 ion beam of H$^-$ in the same apparatus \cite{vergara2021}. The distance between the gas cell 
 and the central CEM was modified between two experimental campaigns with a difference 
 in aspect ratios of 18\%, resulting in no measurable difference between the two 
 sets of CS.

 To verify similar detection efficiencies of both CEMs, the counting rate in the central 
 CEM was measured with the AEF set to zero, corresponding to a total count rate of 
 residual neutral atoms plus the O$^-$ parent ion-beam counts. This counting rate was
 compared with the counting rates of the lateral CEM added to the central CEM's with 
 the AEF on, confirming the total central CEM's count rate. This test was performed
 at each energy, under empty gas cell conditions. It is pointed out that this test also 
 helps to reject any measurable electric field-induced electron detachment by the AEF
 since the total amount of particles remains unchanged with AEF on. The systematic error 
 of the cross-sections has been estimated to be 11\% \cite{hernandez2014}.


 The measurement methods for the collision-induced total electron-loss cross section (CS) 
 are based on the solutions to the fraction $F$ equations for the three possible final 
 projectile's charge states \cite{mcdaniel1993}. For instance, for the neutral atoms $F$ 
 is derived as, $F_0 = (I_0 - I_b)/I_i$ where $I_0$ is the signal count rate of oxygen 
 atoms that resulted from the interaction with the gas, $I_b$ is the ion-beam background 
 count rate and $I_i$ is the count rate of the initial parent ion beam with an empty gas cell.

 Under the present experimental single-collision conditions (see Fig. \ref{fig:scc}) 
 and,  with an ion beam composed of O$^-$ anions, equations for the relevant fractions reduce to: 

 \begin{figure}
 \begin{center}
 \includegraphics[width=\columnwidth]{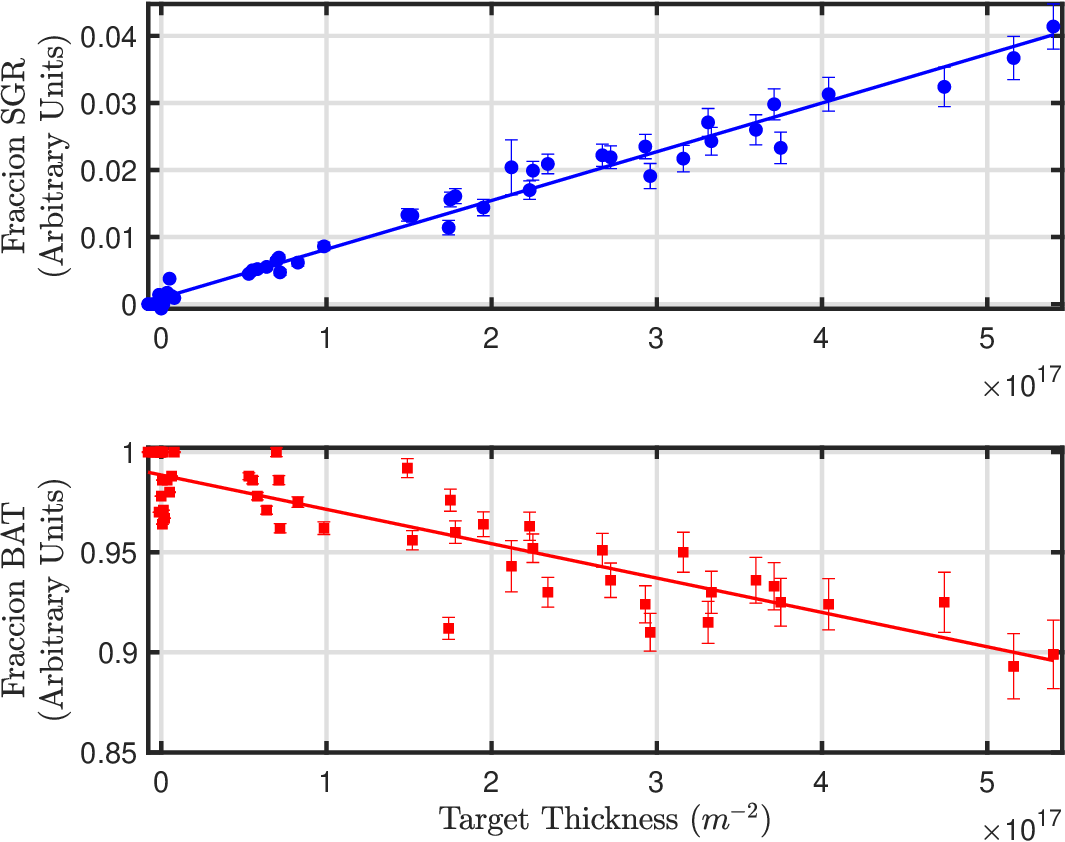}
\caption{\label{fig:scc} 
Examples of the measured data at 2.5 keV. Relevant reduced signals
of the SGR and BAT methods as a function of the target thickness. The top panel corresponds 
to SGR method and the lower panel corresponds to the BAT method. The electron
detachment cross-section is given, at first order, by the slope for each curve.}
 \end{center}  
 \end{figure} 

\begin{align}
 dF_0 & = F_{-1}\sigma_{-10} d\vartheta \label{eqF0} \\
 dF_{-1} & = -F_{-1}(\sigma_{-10} + \sigma_{-11}) d\vartheta \label{eqF-1}  \\
 dF_{1} & = F_{-1}\sigma_{-11} d\vartheta \label{eqF1}
\end{align} 

 Double electron loss cross-section ($\sigma_{-11}$) was estimated by monitoring 
 the yield of O$^+$ by reversing the AEF polarity confirming to be negligible, as it
 is the case for other negative species at this energy \cite{hernandez2016}. 

 Using the signal from the neutral atoms (central CEM), we applied the SGR method, 
 Eq. (\ref{eqF0}). From the ion beam signal (lateral CEM), we applied the BAT method, 
 Eq. (\ref{eqF-1}). The experiment is carried out as a function of the O$_2$ gas 
 target thickness $\vartheta = \ell P/ \kappa T$. Where $\ell$ is its effective 
 length, $P$ the pressure, $T$ the temperature of the gas cell and, $\kappa$ is 
 Boltzmann's constant. Although both measurement methods are expected to yield the 
 same value for $\sigma_{-10}$, we use different notations: $^s\sigma$ for SGR
 and, $^b\sigma$ for BAT, respectively.


\section{Results}

\begin{figure}
\begin{center}
 \includegraphics[width=\columnwidth]{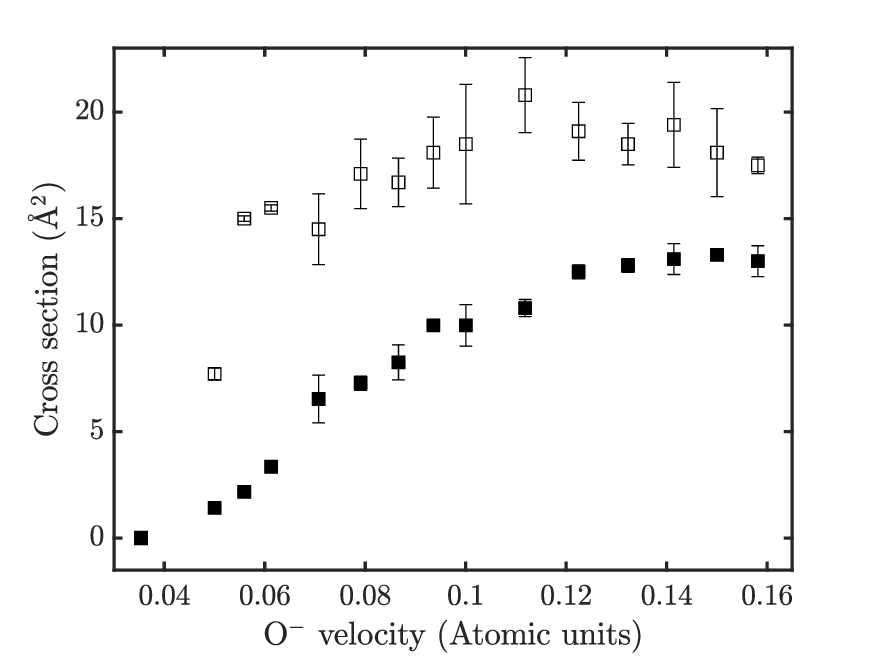}
  \caption{ \label{OurResults} 
  Present results for the measurements of the collision-induced total electron loss cross sections (CS) 
  for the interaction system: O$^-$ + O$_2$. Measured with BAT method (open squares) were 
  derived from experimental parameters in Eq. (\ref{eqF-1}). Measured with SGR method (closed 
  squares) were derived from Eq. (\ref{eqF0}).}
\end{center}  
\end{figure}

 The present measurements are shown in Fig. \ref{OurResults}. The data dispersion in 
 Fig. \ref{OurResults} is mainly originated by the pressure measurements and the ion 
 beam instability caused by the ion source's plasma fluctuations. In the case of $^b\sigma$, 
 the vertical axis is proportional to the ion-beam intensity readings normalized to 
 $I_i$. In the case of $^s\sigma$, the vertical axis is proportional to the initial and 
 final $I_i$ ion-beam currents average. This could be the reason for the difference in 
 the data dispersion.
 
 Below a kinetic energy $E_0$ $<$ 4 keV, CS is lower; consequently, the counting 
 rate signal intensity deteriorated. This effect was averted by increasing the gas 
 cell pressure. This strategy was unfeasible at the higher energy interval because 
 CS was higher and, the signal saturated the counters. This resulted in two 
 different regimes for the statistical error bars in the measurements. 


 We now turn to the question of the disagreement between $^s\sigma$ and $^b\sigma$.
 In BAT, all processes that can extinguish the parent ion beam intensity are accounted 
 for, while in SGR, only neutral atoms resulting from the interaction with O$_2$ are 
 detected. Although, at first order of the solutions of Eq. (\ref{eqF-1}) and Eq. (\ref{eqF0})
 the values for $^s\sigma$ and $^b\sigma$ are expected to be the same, the physics involved 
 in each of the measurement methods is different.
 For instance, double electron loss will not be accounted for in SGR, while it will contribute 
 to BAT. Another process that contributes to CSs is electron transfer, however,
 this process 
 is expected to contribute equally to both signals, therefore, it does not explain 
 the discrepancy.
 
\begin{figure}
    \begin{center}
        \includegraphics[width=\columnwidth]{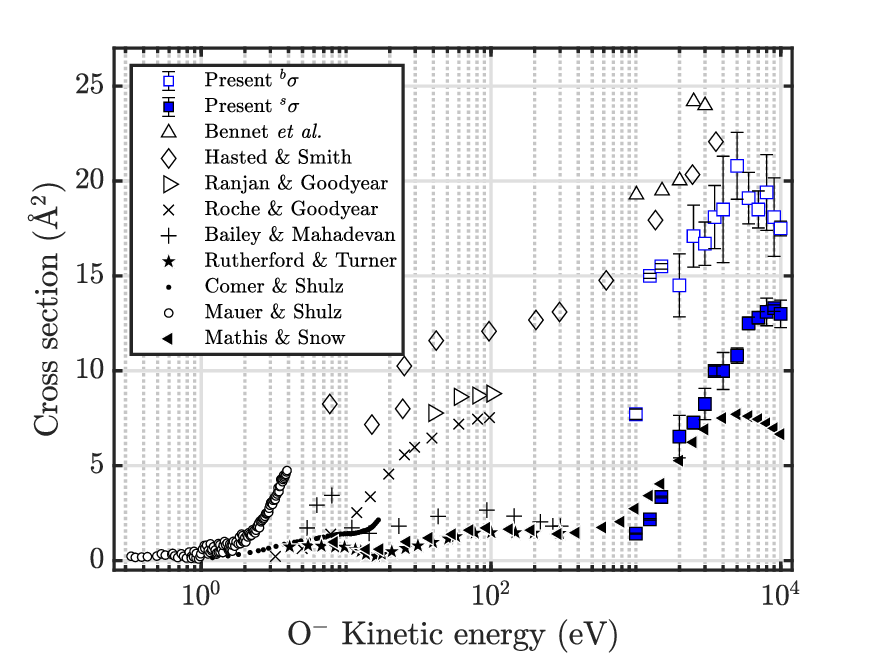}
        \caption{ \label{all-cross-sec} 
        Partial compilation of electron loss cross sections for the interaction system: 
        O$^-$ + O$_2$ from relevant authors. Present results: BAT ($^b\sigma$), SGR ($^s\sigma$).
        Bailey \& Mahadevan \cite{bailey1970}, 
        Hasted \& Smith \cite{hasted1956},  
        Bennet {\it et al.} \cite{bennet1975}.
        Ranjan \& Goodyear \cite{ranjan1973}, 
        Comer \& Shulz \cite{comer1974}, 
        Mauer \& Shulz \cite{mauer1973}.
        Electron transfer cross sections: 
        Mathis \& Snow \cite{mathis1974}
        Roche \& Goodyear \cite{roche1969},
        Rutherford \& Turner \cite{rutherford1967}.
        }
    \end{center}  
\end{figure}

 It is important to point out that the disagreement between the electron loss cross sections
 at the present energy range has been a long-standing question in the field and is not 
 restricted to the present collision system \cite{Rahman1986}. For instance, a partial review 
 of previous measurements for the same collision system is offered in Fig. \ref{all-cross-sec}. 
 Data points measured with BAT (open symbols) are those by Bailey \& Mahadevan \cite{bailey1970}, 
 Hasted \& Smith \cite{hasted1956} and, Bennet {\it et al.} \cite{bennet1975}. In the cases 
 of Ranjan \& Goodyear \cite{ranjan1973}, Comer \& Schulz \cite{comer1974} and Mauer \& Schulz 
 \cite{mauer1973} they measured with techniques based on electron current collection
 with different calibrations and electron collection schemes. 

 In these electron current collection
 techniques an electric or magnetic field is adjusted for maximum current
 collection and a calibration for the electron-collecting electrodes is performed. No
 tendency can be established from these data sets, apart from the observation that since 
 they intended to measure the total electron current; their data can be compared, in principle,
 with BAT.
 
 It is noted in Fig. \ref{OurResults} that the cross-section difference tends to decrease 
 as a function of the particles' speed. We propose that the observed difference in CS 
 can be attributed to the presence of an undetermined auto-detaching state formed 
 during the collision. 
 
 Let's define the time of flight $t_{of}$ as the time the ion beam spends through the length 
 of AEF; this time is different for each accelerating energy. At higher energies,
 $t_{of}$ becomes shorter, and the auto detaching states 
 states have less time to decay. This is the reason
 why at the higher energy interval of the present study, $^b\sigma$ - $^s\sigma$ appears
 to subside. As a consequence, electrons may be lost during the anion's $t_{of}$ 
 to the lateral CEM while in the AEF region, resulting in O atoms not being detected by 
 the central CEM. This explains why they are accounted for by BAT and not by SGR. 

 To try to test this hypothesis, we offer the following analysis. 
 Let the flux of particles per unit area that the ion beam encounters at speed $v$ 
 in the gas cell be $v\vartheta$. Let the AD states flux be proportional to the difference 
 of the total flux responsible for the attenuation and the flux related to direct 
 detachment. This difference decreases as a function of time:
 
\begin{equation}\label{fluxT}
 \int_{0}^{\infty} v \ \vartheta^m dt = - \int_{0}^{\infty} \left( \vartheta^b - \vartheta^s \right) v \ dt
\end{equation}
 
 \noindent The superscripts $m, \ b$ and $s$ refer to metastable, BAT, or SGR techniques
 respectively. At $t$ \ $\rightarrow$ \ $\infty$, $^mF$ \ $\approx$ \ $^bF$ \ $\approx$ \ 
 $^sF$ and  Eq. (\ref{fluxT}) reduced to,
 
 \begin{equation}\label{betapar}
  ^m\sigma = \frac{^s\sigma \ ^b\sigma}{^b\sigma - \ ^s\sigma} 
 \end{equation}

\begin{figure}
\begin{center}
 \includegraphics[width=\columnwidth]{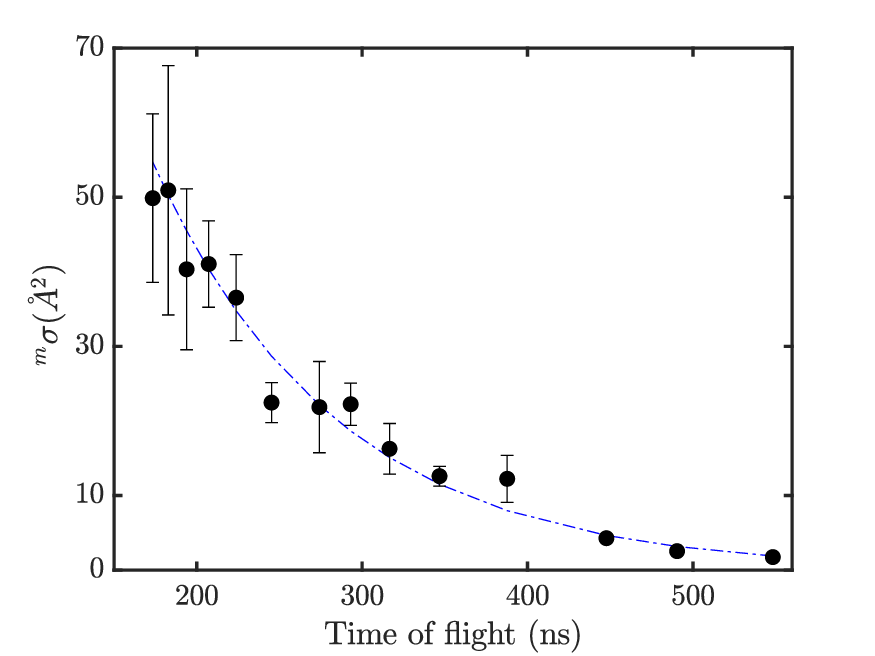}
  \caption{ \label{betaOO2} 
  Auto-detaching CS, $^m\sigma$, of the metastable population as a function of the time of flight 
  derived according to Eq. (\ref{betapar}) for the interaction system O$^-$ + O$_2$ \ 
  $\rightarrow$ O$^*$ + e$^-$ + O$^*_2$. The dashed line represents a minimum square fit 
  to an exponential decay from which a value of $107 \pm 15 \text{ ns}$ for $\tau$ was derived. 
  The error bars represent total uncertainty: the error from the AEF edge effects and the average 
  standard deviation from the numerical fit.}
\end{center}  
\end{figure}

\begin{figure}
\begin{center}
 \includegraphics[width=\columnwidth]{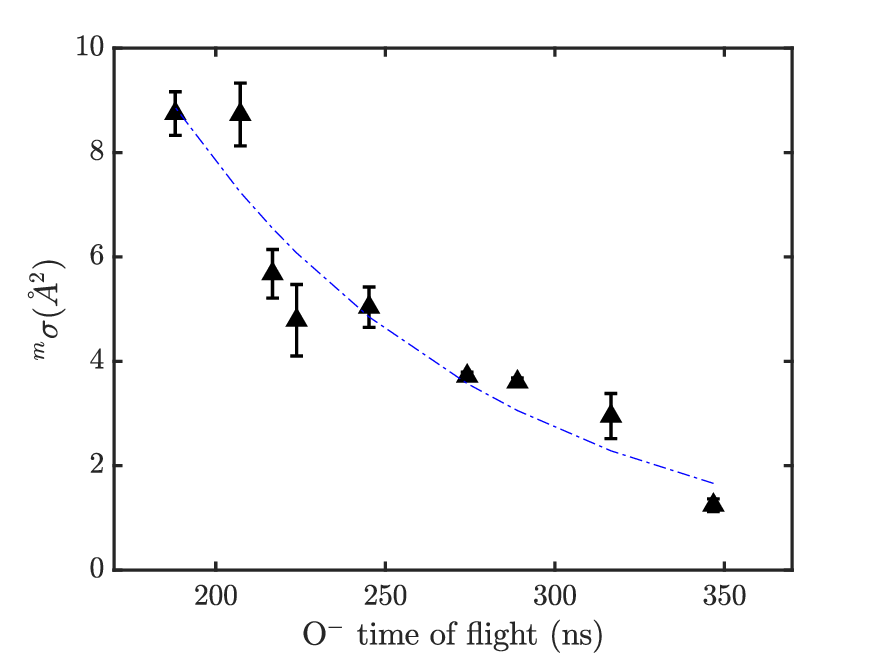}
  \caption{ \label{betaON2} 
  Auto-detaching CS, $^m\sigma$, of the metastable population as a function of the time of flight 
  derived according to Eq. (\ref{betapar}) for the interaction system O$^-$ + N$_2$ \ 
  $\rightarrow$ O$^*$ + e$^-$ + N$^*_2$. The dashed line represents a minimum square fit 
  to an exponential decay from which a value of $95 \pm 13 \text{ ns}$ for $\tau$ was derived. 
  The error bars represent total uncertainty: the error from the AEF edge effects and the average 
  standard deviation from the numerical fit. The data for the cross sections ($^b\sigma$ and 
  $^s\sigma$) were taken from \cite{AAmartinez2024}.}
\end{center}  
\end{figure}

 Plotting $^m\sigma$ versus t$_{of}$, one obtains Figs. \ref{betaOO2} and \ref{betaON2} 
 that show behaviors consistent with the hypothesis of a metastable auto-detaching state 
 population's decay for both gas targets. The cross sections used to derive Fig. \ref{betaON2} 
were taken from \cite{AAmartinez2024}. Both experimental derivations for $^m\sigma$ are
 consistent. We note that the value is lower when the target is N$_2$, possibly due to
 the cross-sections when the target was N$_2$ being lower and, having an impact on 
 the statistical quality of the value. The average for $\tau$ is $100 \pm 10 \text{ ns}$. 

 The auto-detachment lifetime can also be assessed by quantum-mechanical methods as 
 explained above. The detachment width is obtained by analytic continuation of the 
 pseudo-spectrum, but the position of the resonance should be determined to obtain the 
 auto-detachment width and consequently the lifetime. So, a separated calculation was 
 done to determine more precisely the $^4$S (1s$^2$2s$^2$2p$^3$3s$^2$) resonance. This 
 is done with the same approach to obtain the pseudo-spectrum, i.e., at the CASSCF level.
 The calculated energy was 10.83 eV. This value is certainly overestimated 
 since it has been obtained from the pseudo-spectrum, which is not appropriate for this 
 kind of state, {\it i.e.} a state immersed in the continuum, due to the use of 
 square-integrable basis set. Experimentally \cite{spence1975}, this state was determined as 8.78 eV 
 relative to the oxygen ground state. If we sum to that the value of the measured 
 electron affinity of oxygen, 1.46 eV, we end up with 10.24 eV. This should be compared 
 with the present value of 10.83 eV.
 
 The auto-detachment width (and lifetime) are determined to this transition energy. 

\begin{figure}[h!] 
\begin{center}
 \includegraphics[width=\columnwidth]{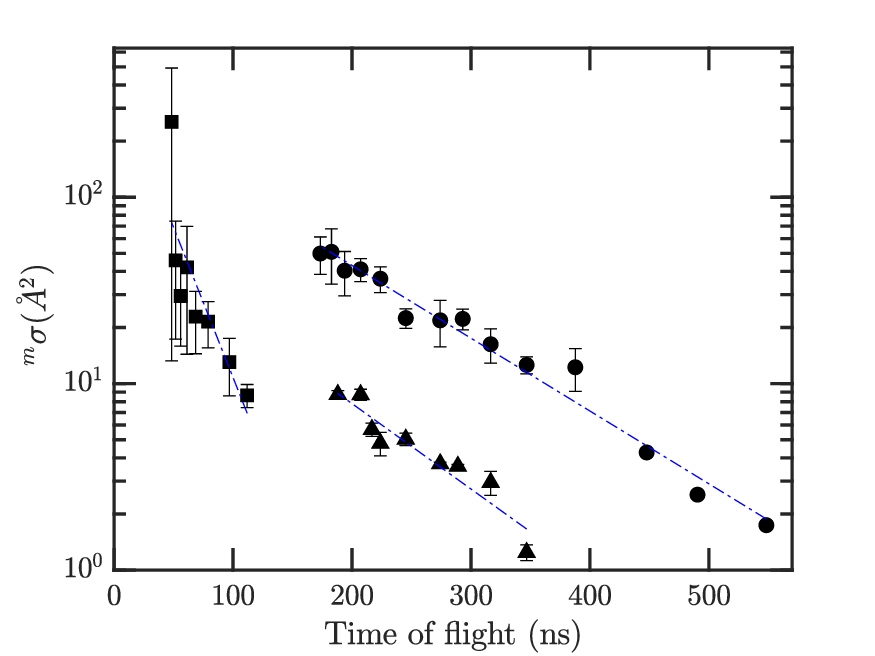}
  \caption{ \label{touslesbetas} 
  For comparison, a compilation plot of the auto-detaching CS, $^m\sigma$, of the metastable 
  population as a function of the time of flight derived according to Eq. (\ref{betapar})
  for the present interaction system: $\bullet$, O$^-$ + O$_2$ (from Fig. \ref{betaOO2}).
  $^m\sigma$ was also derived for O$^-$ + N$_2$: $\blacktriangle$, (from Fig. \ref{betaON2}, 
  data taken from \cite{AAmartinez2024}) and, for H$^-$ + O$_2$, $\blacksquare$ \cite{vergara2021}.
  The dashed lines represent minimum square fits to exponential decays. For the case of
  H$^-$ + O$_2$, $\blacksquare$, $\tau$ = $25 \text{ ns} \pm \text{4 ns}$.
  }
\end{center}  
\end{figure}


\section{Conclusions}

The hypothesis of an auto-detaching state of O$^-$ in the ns time scale is proved
against both experiment and theory. The experiment is based on electron loss cross
sections and the time-of-flight spectrometry principle. The theoretical derivation 
was developed from the Green’s function analytic continuation within the Fano-Feshbach 
model. 
Both methods are consistent with the existence of an undetermined population of a metastable 
auto-detaching state decay with a lifetime $\tau$ of $107 \pm 15 \text{ ns}$ (see Fig.
\ref{betaOO2}) and $95 \pm 13 \text{ ns}$ (see Fig. \ref{betaON2}) in the experiment and, 
of 75 ns in the theory, respectively.

The present hypothesis may also help solving the discrepancies among several measurements
of the electron loss cross sections at the present energy range. 


\begin{acknowledgments}
 National Autonomous University of Mexico (UNAM), through the Support Program for the 
 Improvement of the Academic Staff of UNAM (PASPA). The University of Saskatchewan for 
 the facilities provided during GH sabbatical stay. CONAHCyT CF-2023-I-918. Technical 
 support from Guillermo Bustos, H\'{e}ctor  H. Hinojosa, Reyes Garc\'{i}a, Armando Bustos, 
 Juana A. Romero, and Arturo Quintero. This study was supported in part by the Brazilian 
 agency Funda\c{c}\~{a}o de Amparo \`{a} Pesquisa do Estado do Rio de Janeiro (FAPERJ), 
 code E-262109342019. The authors also acknowledge Conselho Nacional de Desenvolvimento 
 Científico e Tecnológico (CNPq), Coordenação de Aperfeiçoamento Pessoal de Nível Superior 
 (CAPES).
\end{acknowledgments}

\bibliography{bibliografia} 

\end{document}